\documentclass[preprint]{aastex}

\tighten
\input psfig.tex

\newcommand{\asec}{\hbox to 1pt{}\rlap{$^{\prime\prime}$}.\hbox to 2pt{}}

\slugcomment{Submitted to {\it The Astronomical Journal}}
\shortauthors{Lauer et al.}
\shorttitle{Central Luminosity Density Minima}

\begin{document}

\title{Galaxies with a Central Minimum in Stellar Luminosity Density
\footnote{Based on observations made with the NASA/ESA 
{\it Hubble Space Telescope}, obtained at the Space Telescope Science Institute,
which is operated by the Association of Universities for
Research in Astronomy, Inc., under NASA contract NAS 5-26555. These
observations are associated with GO proposals
\# 5454, 5512, 6099, 6587, and 8683.}}

\author{Tod R. Lauer}
\affil{National Optical Astronomy Observatory\footnote{The National Optical
Astronomy Observatory is operated by AURA, Inc., under cooperative agreement
with the National Science Foundation.},       
P.O. Box 26732, Tucson, AZ 85726}

\author{Karl Gebhardt}
\affil{Department of Astronomy, University of Texas, Austin, Texas 78712}

\author{Douglas Richstone}
\affil{Department of Astronomy, University of Michigan, Ann Arbor, MI 48109}

\author{Scott Tremaine}
\affil{Princeton University Observatory, Peyton Hall, Princeton, NJ 08544}

\author{Ralf Bender}
\affil{Universit\"{a}ts-Sternwarte, Scheinerstra\ss e 1, M\"{u}nchen 81679,
Germany}

\author{Gary Bower}
\affil{National Optical Astronomy Observatory, P.O. Box 26732, Tucson, AZ 85726}

\author{Alan Dressler}
\affil{The Observatories of the Carnegie Institution of Washington,
813 Santa Barbara St., Pasadena, CA 91101}

\author{S. M. Faber}
\affil{UCO/Lick Observatory, Board of Studies in Astronomy and
Astrophysics, University of California, Santa Cruz, California 95064}

\author{Alexei V. Filippenko}
\affil{Department of Astronomy, University of California, Berkeley,
CA 94720-3411}

\author{Richard Green}
\affil{National Optical Astronomy Observatory, P.O. Box 26732, Tucson, AZ 85726}

\author{Carl J. Grillmair}
\affil{SIRTF Science Center, 770 South Wilson Avenue, Pasadena, CA 91125}

\author{Luis C. Ho}
\affil{The Observatories of the Carnegie Institution of Washington,
813 Santa Barbara St., Pasadena, CA 91101}

\author{John Kormendy}
\affil{Department of Astronomy, University of Texas, Austin, Texas 78712}

\author{John Magorrian}
\affil{Department of Physics, University of Durham, Durham, United Kingdom,
DH1 3LE}

\author{Jason Pinkney}
\affil{Department of Astronomy, University of Michigan, Ann Arbor, MI 48109}

\author{S. Laine, Marc Postman, \& Roeland P. van der Marel}
\affil{Space Telescope Science Institute, 3700 San Martin Drive, Baltimore, MD
21218}

\vfill


\begin{abstract}

We used {\it Hubble Space Telescope} WFPC2 images to identify six
early-type galaxies with surface-brightness profiles that {\it decrease}
inward over a limited range of radii near their centers.
The implied luminosity density profiles of these galaxies have local minima
interior to their core break radii.  NGC 3706 harbors a high surface
brightness ring of starlight with radius $\approx 20$ pc.
Its central structure may be related to that in the double-nucleus
galaxies M31 and NGC 4486B.
NGC 4406 and NGC 6876 have nearly flat cores that on close
inspection are centrally depressed.  Colors for both
galaxies imply that this is not due to dust absorption.
The surface brightness distributions of both galaxies are consistent with
stellar tori that are more diffuse than the sharply defined system in NGC 3706.
The remaining three galaxies are the brightest cluster galaxies in A260, A347,
and A3574.  Color information is not available for these objects, 
but they strongly resemble NGC 4406 and NGC 6876 in their cores.
The thin ring in NGC 3706 may have formed dissipatively.
The five other galaxies resemble the endpoints of
some simulations of the merging of two gas-free stellar systems,
each harboring a massive nuclear black hole.
In one version of this scenario, diffuse stellar tori are produced
when stars initially
bound to one black hole are tidally stripped away by the second black hole.
Alternatively, some inward-decreasing surface-brightness profiles may
reflect the ejection of stars from a core during
the hardening of the binary black hole created during the merger.

\end{abstract}


\section{Introduction} 

Early-type galaxies are brightest in their centers and fade into the
background at large radii.  There is no shortage of parametric forms
that describe this smooth progression, but all more or less presume
that the density of stars reaches its maximum in the
center and decreases monotonically outwards.  Over the last decade,
{\it Hubble Space Telescope (HST)} imaging has shown that
galaxy centers nearly always have singular surface brightness profiles
of the form $\Sigma_*(r) \sim r^{-\gamma}$
\citep{crane, k94, f94, l95}.
Low luminosity early-type galaxies, in general, have brightness profiles
that are nearly power laws over several decades in radius
with $\gamma \sim 1$ into the {\it HST} resolution limit; for galaxies
in the Virgo cluster this corresponds to radii of only a few parsecs.
In contrast, the most luminous early-type galaxies have cores, defined by
where the outer power law ``breaks'' or transitions to a shallower
inner cusp --- but even there, $\gamma > 0.$ 
\citet{l95} clearly showed that even ``core galaxies'' had central
{\it density} cusps, $\rho_L\propto r^{-\Gamma},$
with $\Gamma$ significantly greater than zero;
``power-law galaxies'' typically had $\Gamma\sim2.$
\citet{g96} verified the \citet{l95}
conclusions, showing that non-parametric inversion of the surface
brightness profiles ratified the existence of density cusps in
nearly all early-type galaxies imaged with {\it HST.}

Massive nuclear black holes may play a critical role in the origin
and survival of core structure \citep{f97}.
This hypothesis is motivated by the dichotomy
in the central structure of elliptical galaxies.
The low luminosity but dense power-law galaxies will be cannibalized
by the high luminosity but more diffuse core galaxies.
The long-term survival of low-density cores in luminous galaxies
appears to demand moderation of any mergers by the black holes;
the cores should have been filled in long ago without a strong tidal
field to disrupt the in-spiraling nuclei of the power-law galaxies.
The theoretical work of \citet{mnm} verifies the \citet{f97}
argument that the initial creation of a core galaxy results
from the merger of two power-law galaxies, each harboring
a massive nuclear black hole.

Black holes may also be required to explain
the double nuclei of M31 \citep{l93} and NGC 4486B \citep{l96},
two notable exceptions to the rule that stellar density
reaches its maximum at the geometric galaxy center.
The origin of these double nuclei is unknown, but their equilibrium and
stability is most easily understood as a consequence of the massive
black holes believed to reside in their centers \citep{km31,k4486b}.
The two visible nuclei in each galaxy
are probably not distinct stellar systems, but may arise
from a torus of stars bound to the black hole in
a Keplerian potential \citep{trem}.  Again, such tori may
result from a merger in which two massive black holes are brought
together by dynamical friction \citep{hbr}.

In this context, we have searched for other galaxies
with unusual central structures that may shed additional light on the
formation of the central structure in galaxies, with particular attention to
early-type systems exhibiting a central minimum in surface brightness.
We identify six systems culled from a large sample
of galaxies imaged by {\it HST} with starlight
distributions that do not neatly fit into
the schema that cores always have cusps with $\gamma>0.$

\section{Observations and Analysis}

\subsection{The Sample and Observations \label{sec:obs}}

The galaxies presented here were identified by searching through a
heterogeneous collection of {\it HST} WFPC2 images of early-type galaxies.
The goal was to identify systems that had minima in their stellar
volume densities interior to their cores.
The search criterion thus was to select galaxies
that did not appear to be affected by dust absorption, but that
had {\it projected} brightness profiles that decreased inward.
This criterion is actually conservative.
A core with an essentially flat profile ($\gamma\approx0$)
in projection over an extended radial range can also be consistent with
a inwardly decreasing density profile.  Galaxies of this sort that may bear
a closer look include NGCs 1600 \citep{byun, g96}, 4291, 5813, and 5982
\citep{rest}.

Of the galaxy images examined, 51 were observed under {\it HST} programs
GO-5512, 6099, and 6587; these are our programs to characterize
the central structure of nearby early-type galaxies, with
emphasis on spectroscopic searches for massive black holes.
We supplemented this sample with the 15
early-type galaxies with kinematically decoupled cores observed
by \citet{carollo} in program GO-5454.  Overall, 66 galaxies were imaged by
these four programs.  

Although the samples observed in each of
these programs were defined with varying criteria, the observational
parameters are fairly uniform.  In all cases the nucleus of
the galaxy was positioned at the center of the WFPC2 high-resolution
PC1 chip.  Images obtained in the F555W ($V$-band) filter were available for
all galaxies; F814W ($I$-band) images were also available in most cases.
The exposures in each filter were generally limited to a single orbit;
this was usually sufficient to obtain a signal-to-noise ratio
($S/N$) of $\sim100$ per pixel in the galaxy centers.
In program GO-6587 we began to use half-pixel dither steps, allowing
the construction of Nyquist-sampled images with double-sampling \citep{l99}.

All images were deconvolved with 40 iterations
of Lucy-Richardson deconvolution \citep{lucy,rich}.
This method is well-suited to WFPC2 data; \citet{l98} show examples
of the deconvolution of simulated observations and compare deconvolved
profiles of galaxies observed with both WFPC1 and WFPC2.
The centers of the galaxies discussed in detail below
were all well resolved, and the deconvolution corrections were modest.
Further, in all cases, the central reductions of surface brightness
were apparent in the original images; deconvolution improves the
accuracy of the photometry, but is not required to recognize the
morphological features discussed below.

After we began work on this paper, an additional sample
of early-type galaxies became available through the WFPC2 ``snapshot''
program GO-8683 (van der Marel, PI; \citealt{laine}) on the central structure of
brightest cluster galaxies (BCGs).  The BCG sample
consists of the \citet{pl} set of 119 BCGs with $z<0.05;$
it is the most homogeneous set of galaxies in all five programs.
These galaxies are of special interest, given the likelihood that some
galactic cannibalism is still going on at this epoch, delivering
faint cluster galaxies to the centers of the BCGs \citep{l88,f97}.
The BCG snapshot images were only obtained with the F814W filter and PC1,
and have significantly lower $S/N$ than the more nearby galaxies.
Our discussion is based on the 75 BCG images observed as of December 2001.

We discuss the six candidate galaxies with
central minima in their density profiles as follows.

\subsection{NGC 3706 \label{sec:n3706}}

\begin{figure}[thbp]
\plotone{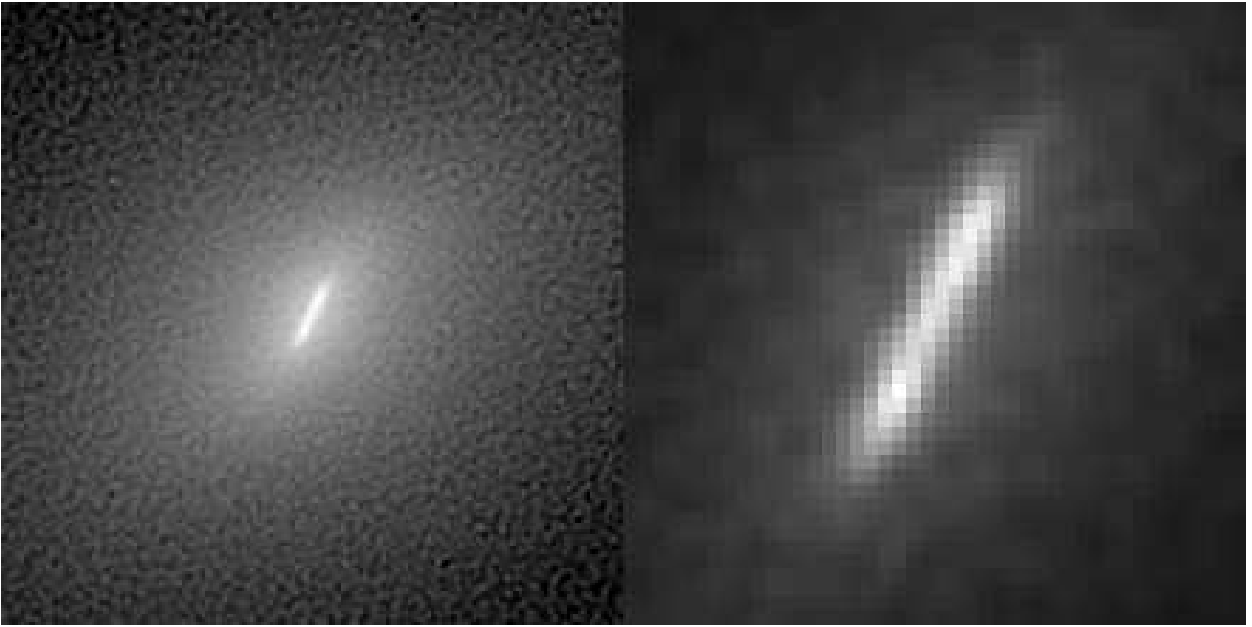}
\caption{The F555W image of NGC 3706.  The panel on the left shows
the central $5''\times 5''$ of the WFPC2 F555W image of NGC 3706.
The galaxy was centered on the PC1 chip.  The image is 
double-sampled, deconvolved, and displayed with a logarithmic stretch.
The right panel is magnified by a factor of 5 to show the ring itself;
the stretch is linear in this image.
North is at $219.0^\circ$ measured counterclockwise from the
vertical axis.}
\label{fig:n3706_im}
\end{figure}

NGC 3706 is an S0 galaxy in the \citet{f89} group 242
(2749 km s$^{-1}$ group velocity).  Its luminosity lies in the
transition zone between power-law and core galaxies \citep{f97}.
\citet{cd} obtained extensive ground-based photometric and spectroscopic
observations of NGC 3706; they found the galaxy to have
strong central rotation and a pronounced velocity dispersion peak.
The WFPC2 images show that NGC 3706 harbors a bright compact
edge-on stellar ring or torus at its center (Figure \ref{fig:n3706_im}).

\begin{figure}[thbp]
\plotone{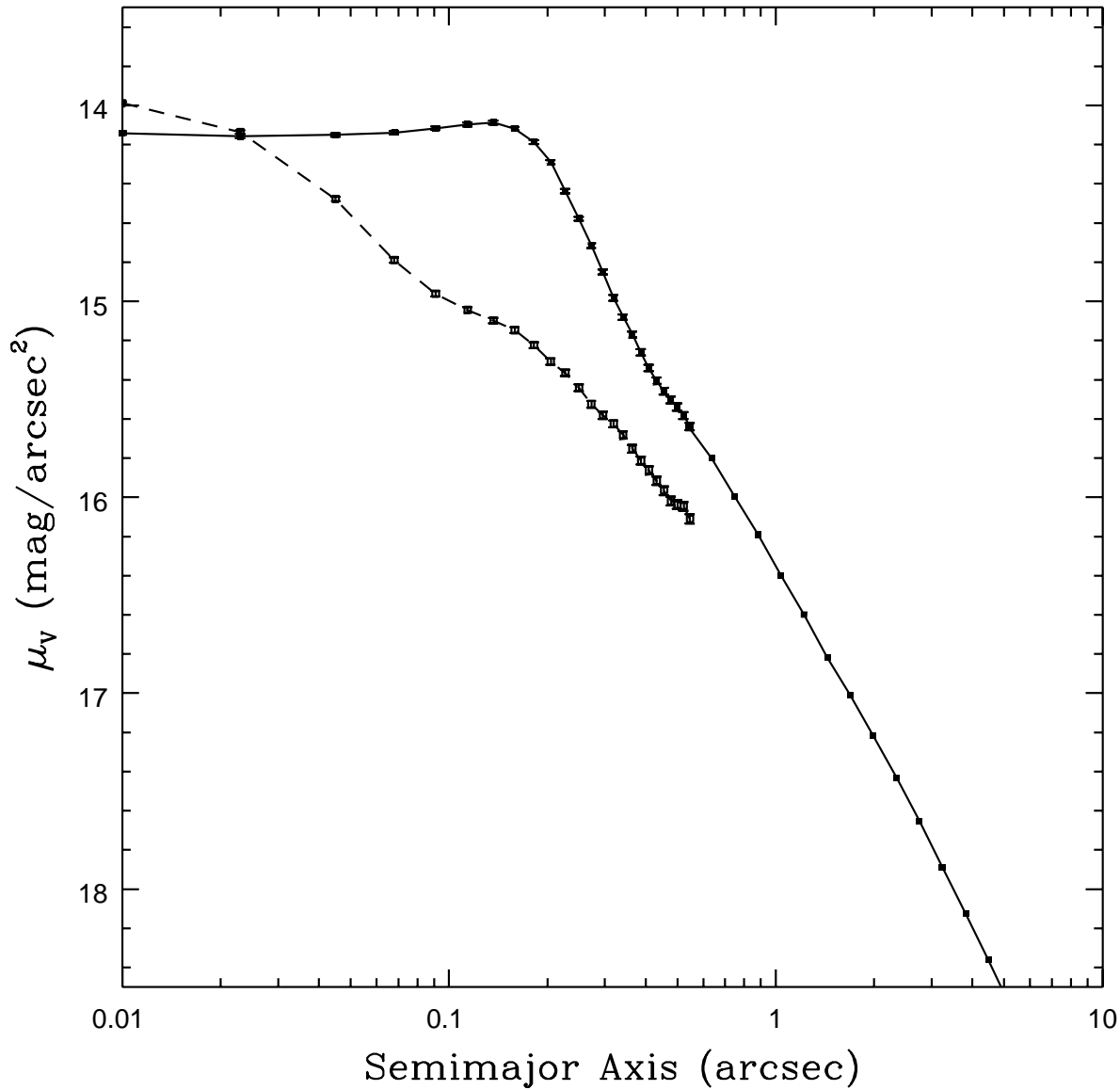}
\caption{Surface brightness profiles for NGC 3706.
The solid line is the major axis brightness profile.
For $r<0\asec5$ the profile is measured from a cut ($0\asec09$
in width) along the disk, with the opposite sides of the nucleus
averaged.  At larger radii, standard isophote fitting is used (connected dots).
The dotted line is a cut perpendicular to the ring, again with
both sides averaged.}
\label{fig:n3706_cut}
\end{figure}

A brightness profile (Figure \ref{fig:n3706_cut}) measured along the major axis
of the ring shows that its surface brightness rises from the center
by 0.07 mag to a local maxima at $0\asec13,$ or 21 pc
from the galaxy center on both sides.\footnote{$H_0=80~{\rm
km~s^{-1}~Mpc^{-1}}$ is adopted throughout the paper.}
This limb brightening
suggests that the structure is not a filled disk.\footnote{With just one
filter, it is not possible to rule out dust absorption as an alternative
explanation for the central dip in the ring brightness.  There is no
morphological sign of dust outside the ring, however; any dust disk
within the stellar disk would necessarily have a more limited radial extent
and a scale height at least as compact as the stellar system.}
The image of NGC 3706 was dithered and thus has subpixels with $0\asec0228$
scale.  The profile shows the intensity at each subpixel along the
major axis averaged over the width of
a slice four subpixels ($0\asec091$) thick centered on the ring,
with each half of the major axis averaged about the center.

\begin{figure}[thbp]
\plotone{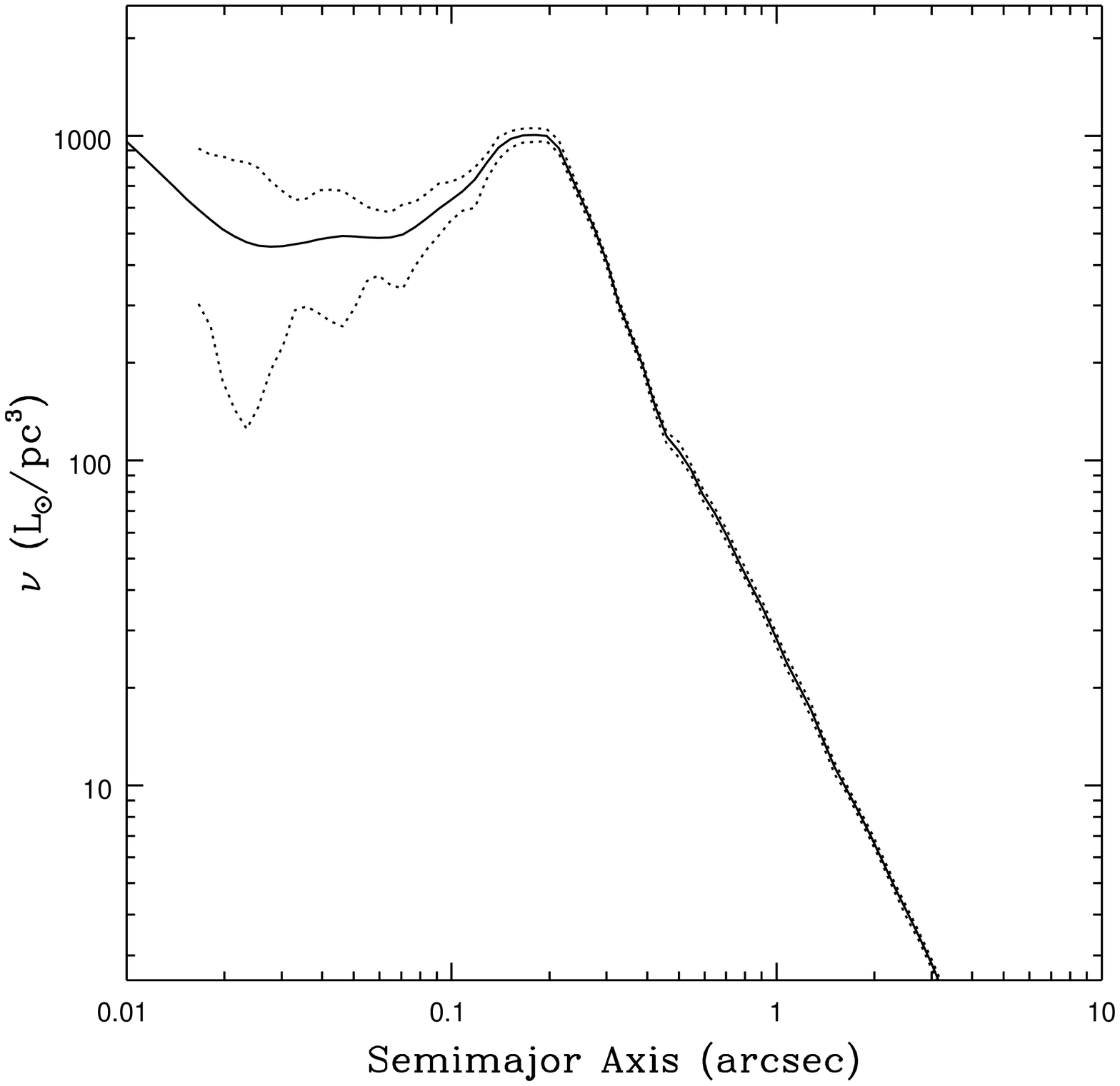}
\caption{$V$-band luminosity density profile for the major axis
of NGC 3706.  Dashed lines give the $\pm1\sigma$ error envelopes.
The profile has been corrected for foreground extinction.}
\label{fig:n3706_den}
\end{figure}

A small inward decrease in surface brightness implies a larger corresponding
decrease in the luminosity density profile,
which is shown in Figure \ref{fig:n3706_den}.
This was computed by performing a non-parametric
Abel inversion (see \citealt{g96}) of the brightness profile
along the apparent ring-plane, assuming that the ring and surrounding
galaxy were axisymmetric within this plane.
The luminosity density of the ring has a sharply defined maximum at
$0\asec18,$ or 31 pc, falling by a factor of $\sim 2$ for $r<0\asec1.$
The small increase in brightness as $r \rightarrow 0$ is of
marginal significance.
A rough estimate of the total ring luminosity is
$\sim 1.3\times10^8L_\odot$ ($V-$band), measured by integrating
the light within a $0\asec68\times0\asec16$ slice centered on the
ring, crudely corrected for the ``background'' galaxy light,
measured in a similar slice slightly offset along the minor axis.

The thickness of the ring appears to be unresolved by {\it HST.}
The minor-axis brightness profile (Figure \ref{fig:n3706_cut})
begins to steepen at $r<0\asec1,$ which is well outside the
resolution limit.  However, measuring the thickness of the ring requires
an assumed light distribution for the galaxy background.
When simple estimates of the background are considered,
it appears that the half-power point of the ring 
vertical extent is less than $0\asec04,$ or 6 pc.

The small size and extreme aspect ratio of the ring in NGC 3706 raise the
issue of the ring's lifetime against thickening by two-body relaxation. From
the stellar velocity dispersion versus black-hole mass relationship of
\citet{g00} and the velocity-dispersion observations of \citet{cd}, NGC 3706
is expected to have a central black hole of mass
$M_\bullet\approx5\times10^8M_\odot.$ This mass would dominate the stellar
mass enclosed within the radius of the ring and implies an angular rotation
rate of $\Omega=2.8\times10^{-13}{\rm~s}^{-1}$ at $r=31$ pc.
If we assume a Gaussian vertical luminosity density distribution of the ring,
$j(R,z)=j_0(R)\exp(-z^2/2z_0^2),$ then the upper limit on the ring thickness
implies $z_0<5 {\rm~pc,}$ which in turn allows calculation of an upper limit
to the vertical velocity dispersion within the ring, $\sigma_z=z_0/\Omega< 43
{\rm~km~s^{-1}}.$ If the shape of the velocity ellipsoid is similar to that in
the solar neighborhood, then the isotropized dispersion is about 1.5 times as
large, or $\sigma<64 {\rm~km~s^{-1}}.$ At the upper limit for the dispersion,
the relaxation time \citep{bnt} within the ring is
\begin{eqnarray}
t_r&=&0.34\sigma^3\left(G^2m\rho\ln\Lambda\right)^{-1} \\
   &=&3\times10^{11}{\rm\ yr}\left(1M_\odot\over
m\right)\left(M\over L\right)^{-1},
\end{eqnarray}
where we have adopted a midplane density
$\rho_0=10^3M_\odot~{\rm pc^{-3}}$ (see Figure \ref{fig:n3706_den}),
the mass-to-light ratio $M/L$ is in solar units, the stars are assumed to have
solar mass, and $\ln\Lambda=\ln(1.2z_0\sigma^2/(GM_\odot))=16.$
This result for the relaxation time is consistent with the hypothesis that the
ring age is comparable to a Hubble time. However, the dependence of $t_r$ on
$\sigma$ and $\rho_0$ implies $t_r\propto z_0^{-4};$ thus $t_r$ may be
substantially shorter than the present estimate, and the actual ring thickness
could be determined by two-body relaxation if $z_0$ were a factor of two
or more smaller than the observational upper limit.

\begin{figure}[thbp]
\plotone{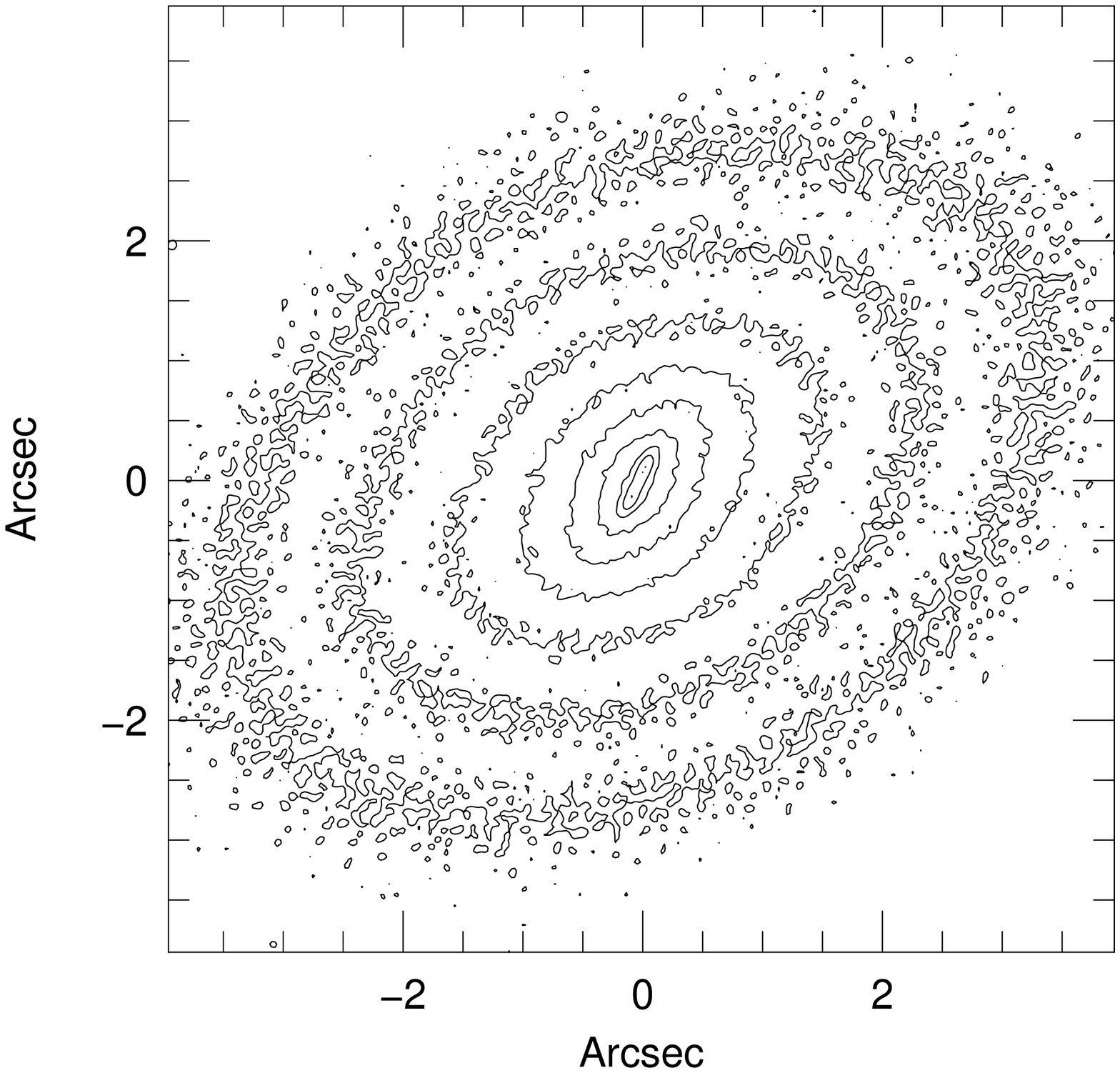}
\caption{A contour map of NGC 3706 in the deconvolved F555W image.
Contours are spaced by 0.5 mag in surface brightness; the outermost
contour corresponds to $\mu_V=18.0$ mag arcsec$^{-2}$.
North is at $219.0^\circ$ measured counterclockwise from the
vertical axis.}
\label{fig:n3706_con}
\end{figure}

\begin{figure}[thbp]
\plotone{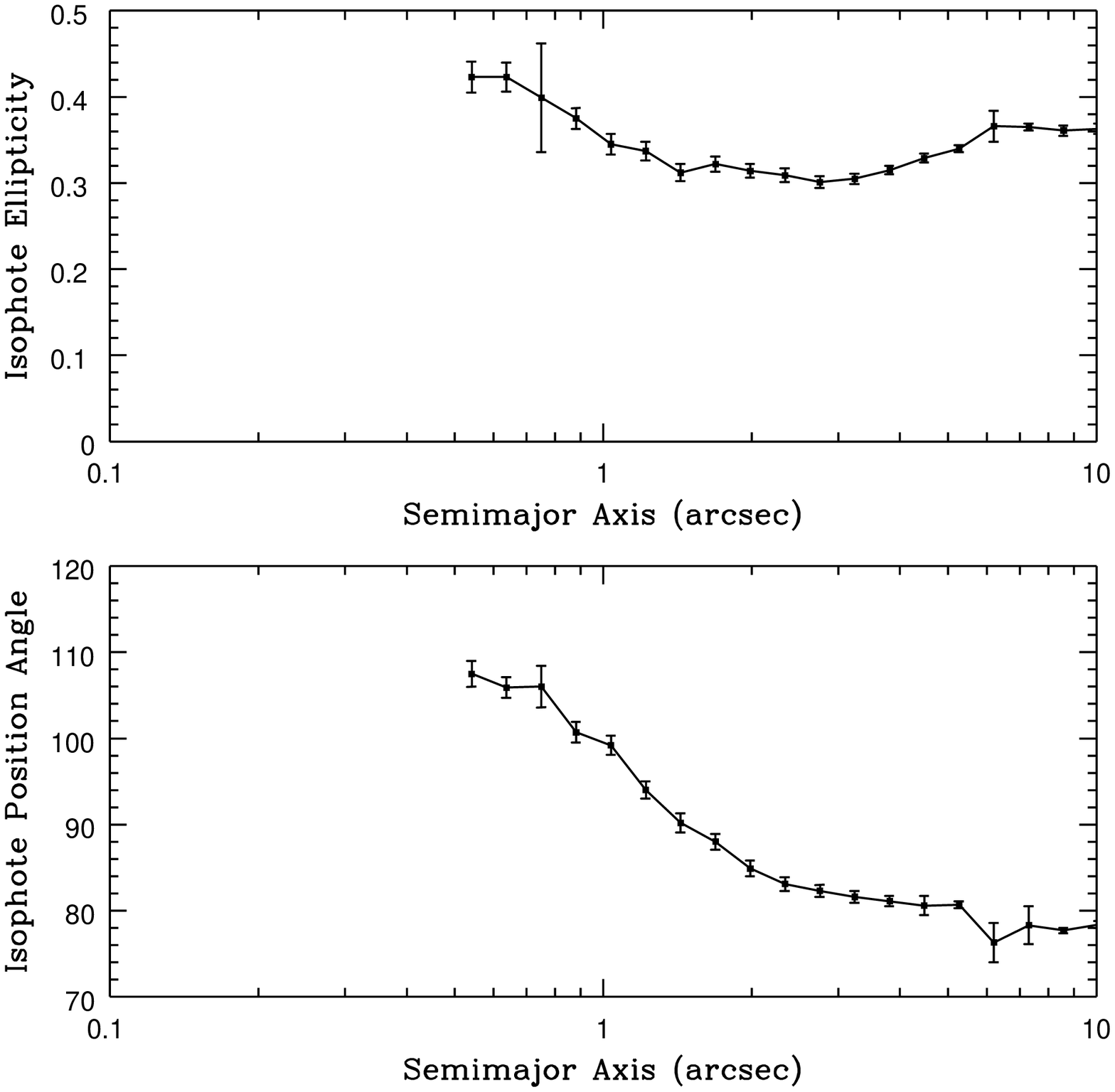}
\caption{Isophote ellipticity and position angle profiles for NGC 3706.
No points are shown for isophotes with semimajor axes $<0\asec5,$
as these isophotes are poorly fitted by ellipses.}
\label{fig:n3706_pa}
\end{figure}

The brightness of the ring falls rapidly from $r \approx 0\asec2$ to
$r \approx 0\asec4,$ where an inflection point in the major axis luminosity
density profile occurs.
The ring is actually misaligned with the major axis of the galaxy
at large radii, as can be seen in the contour map (Figure \ref{fig:n3706_con})
and plot of isophote position angle (PA; Figure \ref{fig:n3706_pa}).
The PA of the ring is $114^\circ,$ as compared
to the galaxy PA of $78^\circ$ for isophotes with $r>5''.$
Even though the bright portion of the ring is compact, the
transition of PA from the ring to outer-galaxy isophote orientation
takes place smoothly over $0\asec5<r<2\asec0;$ this is a strong twist,
given the high ellipticity of the isophotes over the same radii.

The existence of this twist raises an interesting dynamical problem.
If the isodensity surfaces in the galaxy are triaxial
ellipsoids with aligned principal axes, then isophote twists can arise if the
axis ratios of the isodensity surfaces vary with radius and the line of sight
does not lie in one of the principal planes (e.g., \citealt{bm98}).
However, in NGC 3706 we see the ring edge-on. Therefore the isophote
twist between 0\farcs5 and 2\farcs0 implies that the ring plane is not one of
the principal planes of the galaxy at larger radii ($> 300$ pc). Thus
the tidal force from the galaxy must induce precession of the ring plane;
crude estimates suggest that the precession time is a few times $10^8$ yr.
Gravitational interactions between the precessing ring plane and passing
stars from the host galaxy can damp or excite the inclination of the ring
\citep{dk95, nt95}; the damping timescale is
difficult to estimate accurately but typically is only a few precession times
and hence probably is short compared to a Hubble time. The non-zero
inclination might then indicate either that the ring was young, or that the
interactions with the host galaxy have excited the inclination of a ring that
was initially located in or near the equatorial plane of the galaxy.

We stress that it is essentially impossible to decompose
NGC 3706 uniquely into separate background galaxy and ring components;
however, the minor-axis surface brightness profile
does appear to ``break'' at $r \approx 0\asec2,$ which is outside radii
associated with the ring itself.  The slope of the minor axis profile
for $0\asec1<r<0\asec2$ is $\gamma\approx0.4,$ suggestive of a
transition to a shallow cusp.  NGC 3706 thus appears to be a core galaxy.

\subsection{NGC 4406}

NGC 4406 is a giant elliptical galaxy in the Virgo cluster.
NGC 4406 rotates slowly about its {\it major} axis \citep{wagner, franx},
but its core interior to $r<5''$ rotates rapidly about
the minor axis \citep{bender,franx} and has significantly
higher line-strengths compared to the envelope \citep{bs92},
intriguing results in light of the following discussion.
\citet{carollo} obtained F555W and F814W
WFPC2 images of NGC 4406 as part of their sample of elliptical galaxies with
kinematically distinct cores.  They found NGC 4406 to have a well-resolved
core with a ``break radius'' $r_b=0\asec95,$ with a ring of
reduced surface brightness interior to this radius.
Carollo et al. believed that this ``moat'' (our coinage) was due to
dust absorption, but noted that no reddening was seen in a comparison of
$V$ and $I$ images.  We argue instead that the moat is a true reflection
of the starlight distribution in NGC 4406.
One possibility is that NGC 4406 began with a normal
core, and that some of the stars in the region interior to the core break
radius were subsequently ejected; a second is that a diffuse stellar torus is
encircling the nucleus at slightly larger radii, creating the appearance of
reduced surface brightness interior to the torus (this could either be a
face-on torus or a thick edge-on torus; for reasons given below we prefer the
latter interpretation).

\begin{figure}[thbp]
\plotone{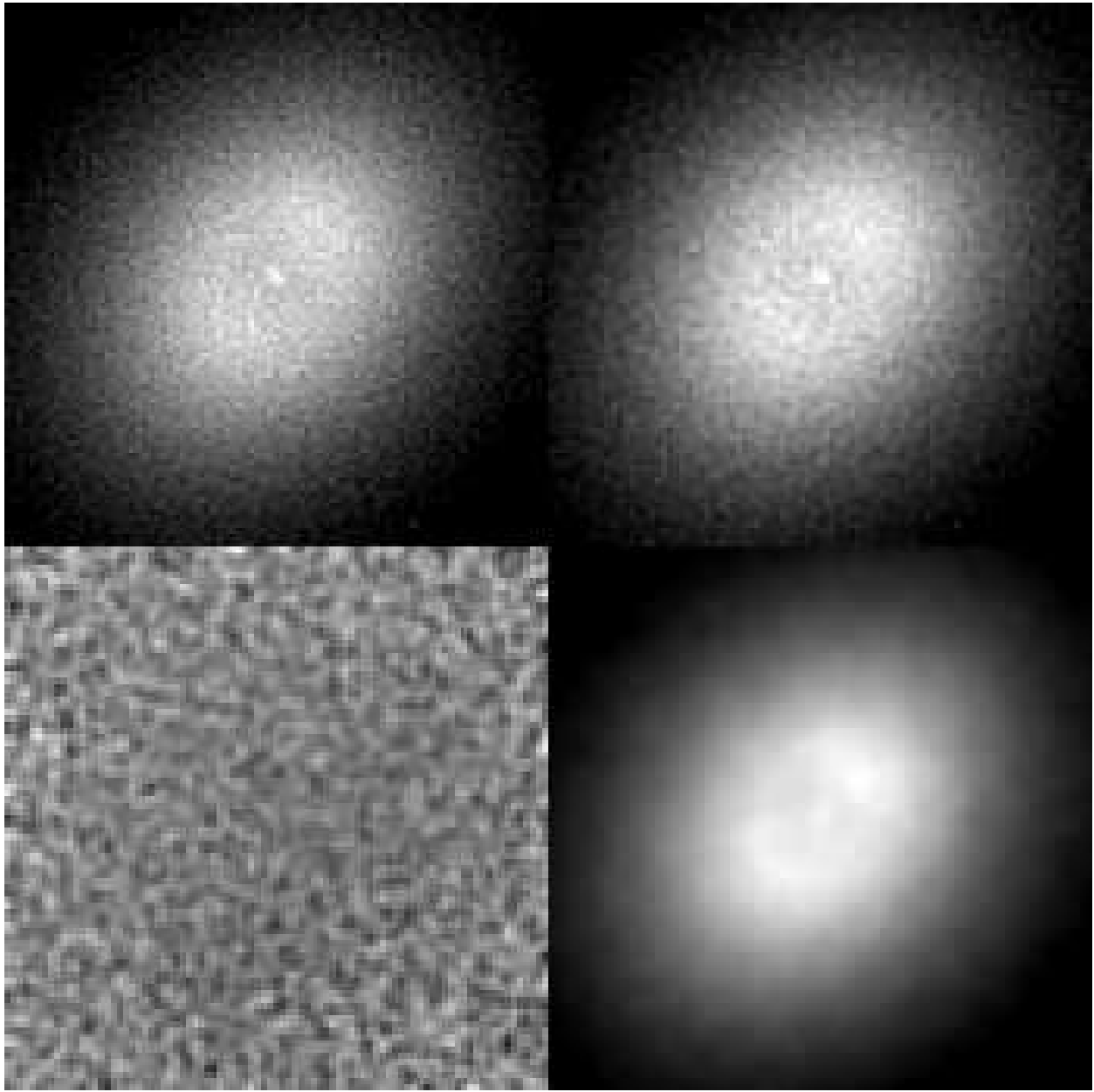}
\caption{Images of NGC 4406 are shown in three different filters.
The upper left and upper right panels are the deconvolved WFPC2 F555W and F814W
images.  The lower right panel is the deconvolved NIC2 F160W image resampled
and rotated to match the WFPC2 images.  The area of the panels is
$4'' \times 4''.$  An arbitrary linear stretch (the same in all three panels)
has been used to enhance contrast in the core.  The lower left panel
is a slightly smoothed F555W/F814W ratio image; the full range of its
gray scale is $\pm10\%.$  North is $8.9^\circ$ measured counterclockwise
from the vertical axis. The nuclear point-source is most evident in the
$V$-band image.}
\label{fig:n4406_im}
\end{figure}

\begin{figure}[thbp]
\plotone{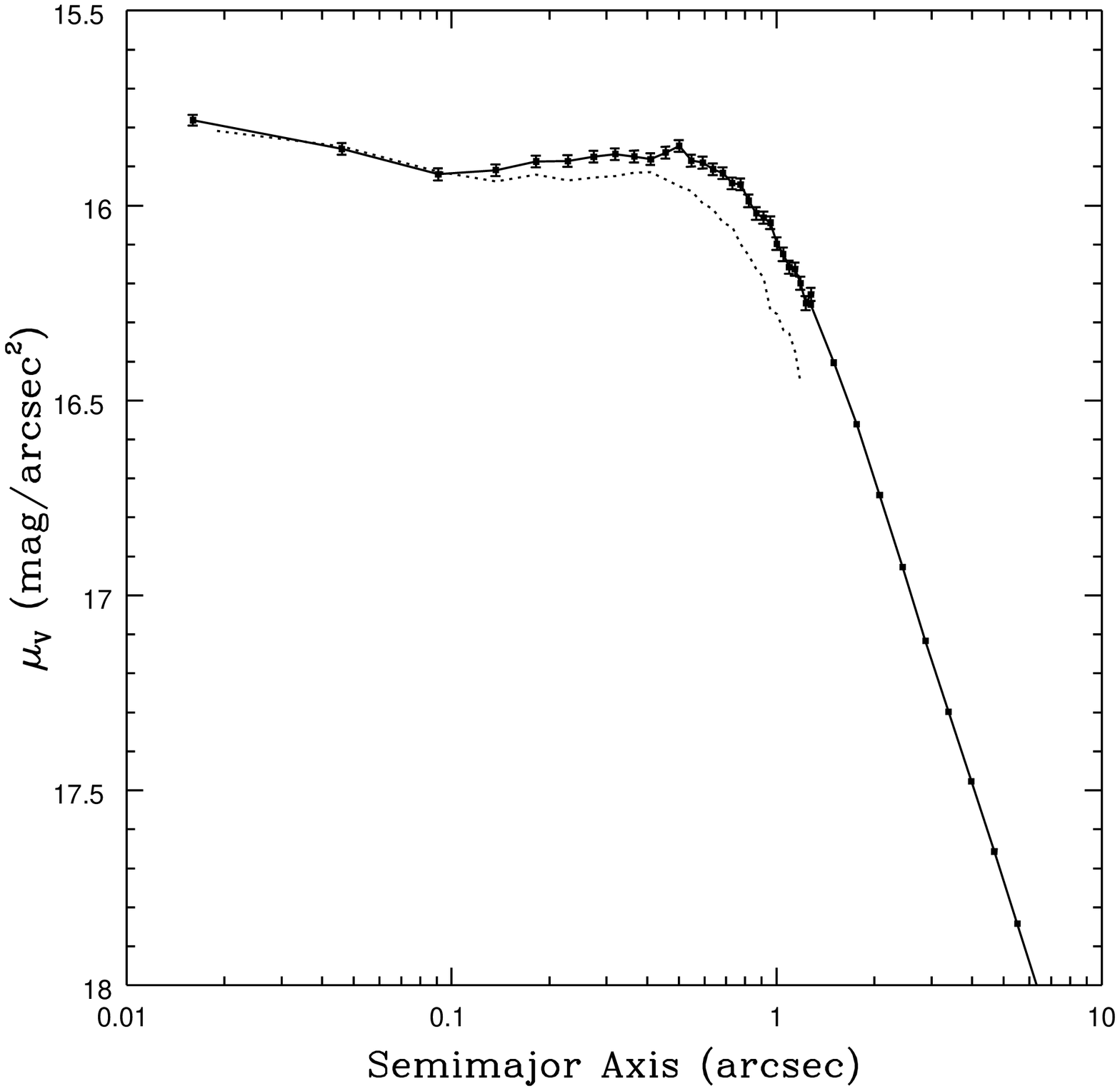}
\caption{Major axis (solid) and minor axis (dotted) deconvolved F555W
brightness profile for NGC 4406.  The profile at $r>1\asec2$
was measured by isophote fitting (connected dots).  At smaller radii the profile
for each axis was measured from a cut along the axis (of width 0\farcs14),
with the opposite sides across the nucleus averaged.}
\label{fig:n4406_cutv}
\end{figure}

Figure \ref{fig:n4406_im} shows the deconvolved
WFPC2 F555W and F814W images of NGC 4406, as well as an
STScI {\it HST} archive F160W NIC2
image of the same region obtained by \citet{rav}.
A moat of reduced emission is visible in all three images.
Figure \ref{fig:n4406_cutv} shows the F555W major axis and minor-axis
surface brightness profiles of NGC 4406.  For $r>1\asec2,$ the profiles were
derived by the standard fitting of ellipses to the isophotes; at smaller radii,
where the isophotes were poorly fitted by ellipses, the profiles consist
simply of cuts along each axis (of 0\farcs14 width), with the opposite
sides about the nucleus averaged. The moat of reduced emission is visible
as a local minimum in the major axis brightness profile (PA$=127^\circ$)
at $0\asec11$ from the nucleus. The profile then brightens outward by 0.07 mag
to a maximum at $0\asec49$ from the nucleus.  The minor-axis profile
is somewhat flatter over the same radii, with the surface brightness rising
only 0.02 mag outward to a maximum at $0\asec38$ (29 pc) from the nucleus.
Both profiles show a sharp peak at the smallest radii,
which appears to be a poorly-resolved nuclear point source.

\begin{figure}[thbp]
\plotone{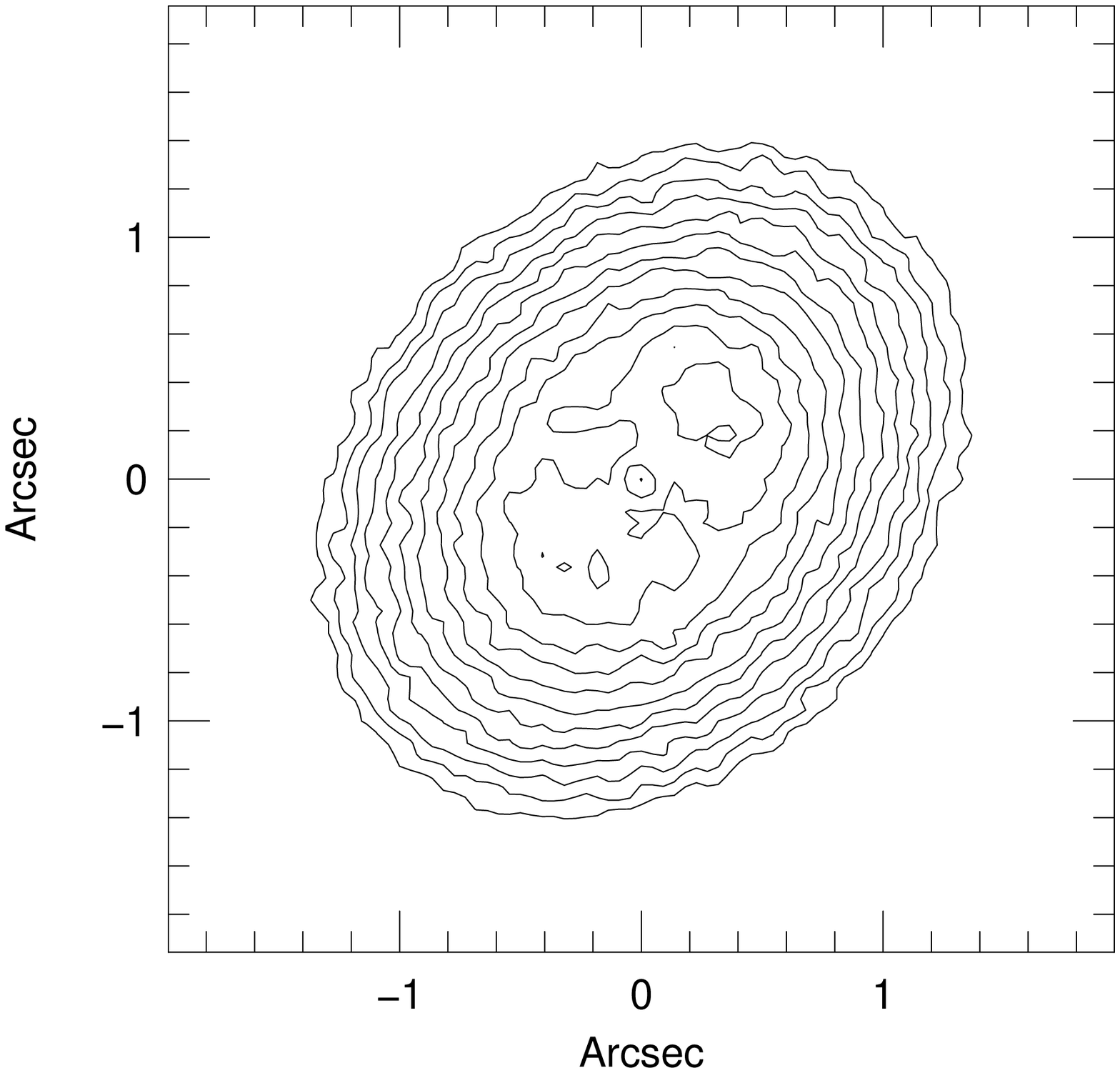}
\caption{A contour map of NGC 4406 in the deconvolved F555W image.
Contours are spaced by 0.05 mag in surface brightness; the outermost
contour corresponds to $\mu_V=16.42$ mag arcsec$^{-2}$. North is 
$8.9^\circ$ measured counterclockwise
from the vertical axis.}
\label{fig:n4406_con}
\end{figure}

\begin{figure}[thbp]
\plotone{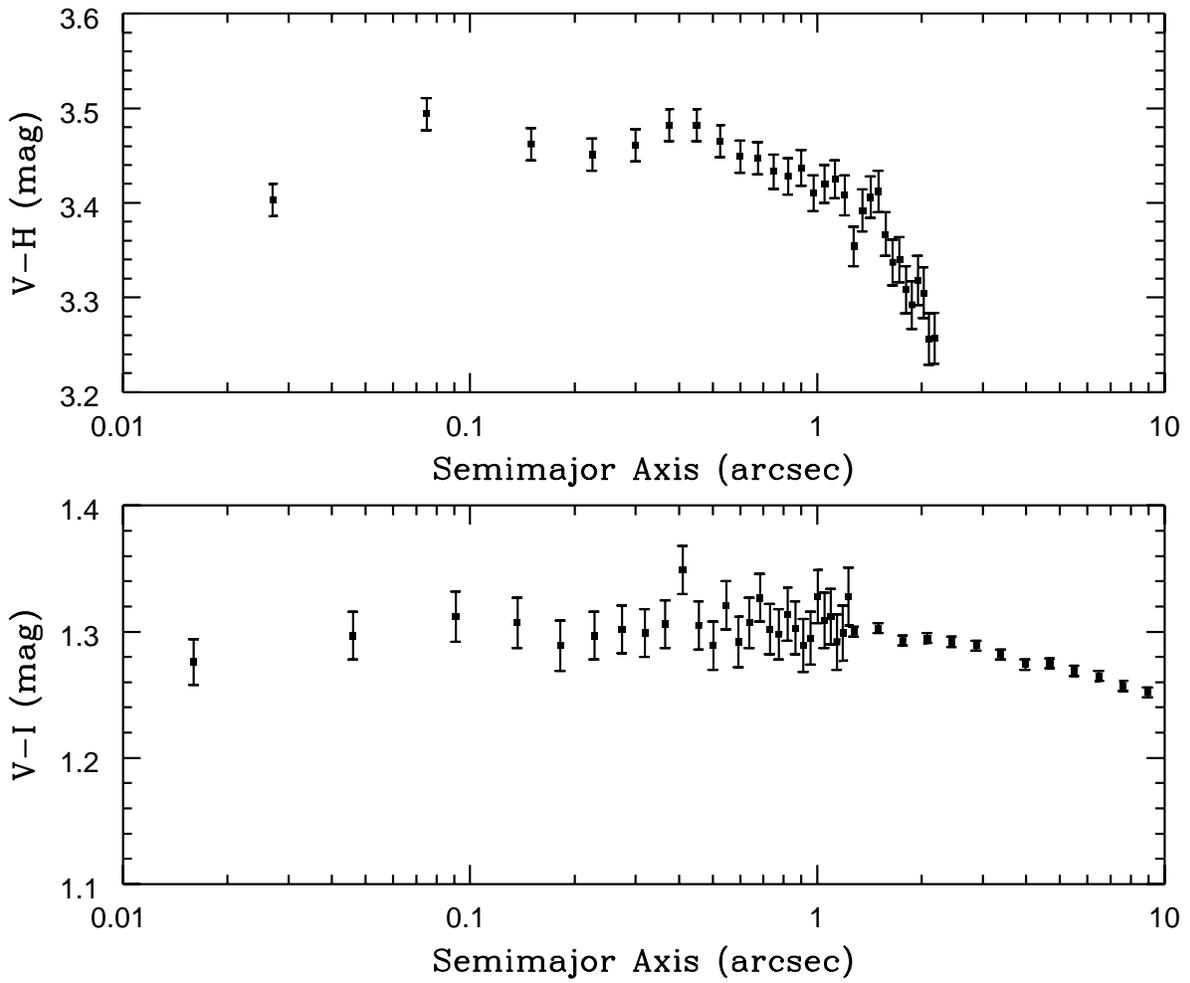}
\caption{Major axis $V-H$ and $V-I$ color profiles of NGC 4406.}
\label{fig:n4406_col}
\end{figure}

\citet{carollo} noted the somewhat lower maximum brightness along the minor
axis and argued that their ``dust ring'' was elongated in this direction.
An equivalent picture, however, is that we are looking at a diffuse, edge-on,
thick torus of stars that in projection is elongated along
the major axis; this interpretation is made somewhat more explicit
by the contour map presented in Figure \ref{fig:n4406_con}.
In this case it is natural to assume
that both the background galaxy and the torus are axisymmetric with respect to
the projected minor axis of the galaxy, and symmetric with respect to the
plane formed by the line of sight and the projected major axis.
We are motivated to consider a dustless interpretation both by the example of a
stellar torus in NGC 3706, and by the lack of any dust reddening in NGC 4406.
Figure \ref{fig:n4406_im} shows that the $V-I$ color map made by dividing
the F555W by the F814W image is devoid of any structure or gradients
over the extent of the core.  NGC 4406 does have a red
color gradient as $r \rightarrow 0,$ with $\Delta(V-I)/\Delta\log(r)=-0.062$ for
$1''<r<10''$ (Figure \ref{fig:n4406_col}), but this color gradient
is completely normal for giant elliptical galaxies \citep{carollo}.
The color for $r<1''$ actually appears to be
constant with $V-I=1.31\pm0.01$ mag.
The color of the moat, itself, measured from all pixels falling
within the annulus $0\asec13<r<0\asec26$ is $V-I=1.309\pm0.006$ mag,
as compared to $V-I=1.303\pm0.006$ mag measured in two $0\asec32$ square
patches centered on the points of maximum surface brightness on
each side of the nucleus.
From the extinction tables in \citet{holtz}, we infer
$A_V\approx2.5\Delta(V-I)$ for the WFPC2 filter-set.
The difference $\Delta(V-I)=0.006\pm0.009$ mag is insignificant,
and the implied $A_V$ is clearly insufficient to account for the
moat of reduced surface brightness.

The case against dust absorption is made particularly strong by the close
similarity of the F160W NIC2 image to the WFPC2 images (information
that was not available to \citealt{carollo}).  Again, the $V-H$
color (Figure \ref{fig:n4406_col}) is essentially flat over the core interior;
if anything, the $V-H$ may become slightly {\it bluer} interior to the
radius of maximum brightness in the core.

\begin{figure}[thbp]
\plotone{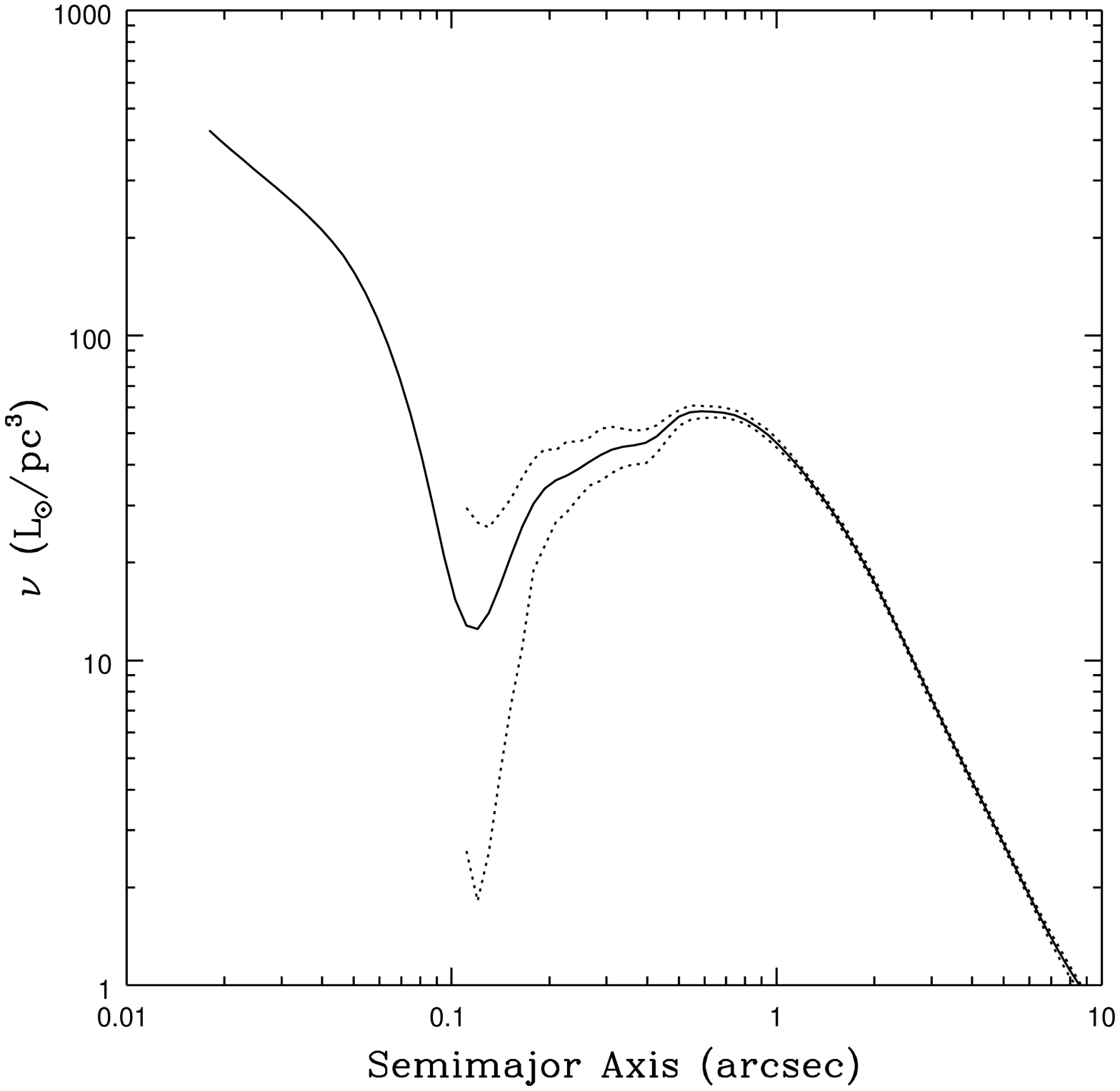}
\caption{$V$-band luminosity density profile for the major axis
of NGC 4406.  Dashed lines give the $\pm1\sigma$ error envelopes.
The profile has been corrected for foreground extinction.}
\label{fig:n4406_den}
\end{figure}

If the unusual core of NGC 4406 reflects the
intrinsic distribution of starlight, then a ring of
even slightly reduced surface brightness implies a stronger corresponding
decrease in luminosity density.  Figure \ref{fig:n4406_den}
shows the implied major axis luminosity density profile
obtained from nonparametric Abel inversions of the profiles shown in Figure
\ref{fig:n4406_cutv}, again done under the assumption that the galaxy is
axisymmetric in the plane defined by the major axis and line-of-sight.
The 0.07 mag inward decrease in surface brightness
on the major axis corresponds to a factor of 2 to 30 decrease
in luminosity density over the same radii.
The exact density profile is clearly sensitive to small
changes in the brightness profile; however, we note that even
for a perfectly flat light profile, NGC 4406's core is sharp
enough ($\gamma=0$ but $\alpha>2$ in the parametric form of \citealt{l95})
that a inward decrease in volume density would still be implied.

Is is ambiguous as to whether the NGC 4406 density profile has been created
by either adding stars to or removing them from an originally normal core.
Fits to the $V$-band profile give $I_b=16.03$ mag arcsec$^{-2}$ 
and $r_b=0\asec93,$ corresponding to 72 pc at an assumed
distance of 16 Mpc.  This implies that NGC 4406 has a normal core
for its luminosity \citep{f97}.  There is no unique way
to decompose the density profile into a torus superimposed
on a more normal density profile.  Under some assumptions,
such an exercise could imply that NGC 4406 had a substantially
larger core prior to some sort of accretion event;
scatter in the $L-r_b$ and $L-I_b$ relationships is too large
to rule this out \citep{f97}.
Conversely, NGC 4406 may have had a {\it smaller} core prior to the
event --- indeed, it is within the luminosity range of the
power-law galaxies, which have no cores at all.
One could thus argue that the present density profile represents
an evacuation of a plausibly more concentrated initial profile.

\subsection{NGC 6876}

\begin{figure}[thbp]
\plotone{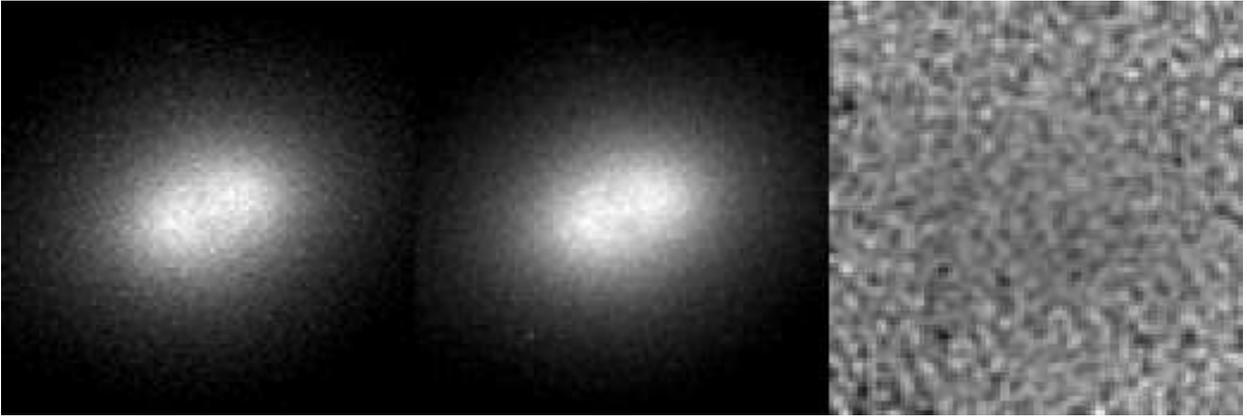}
\caption{Images of the NGC 6876 core are shown in two filters.
The left and middle panels are the deconvolved WFPC2 F555W and F814W images.
The area of the panels is $4'' \times 4''.$  An arbitrary linear stretch
has been used to enhance contrast in the core.  The right panel
is a slightly smoothed F555W/F814W ratio image; the full range of its
gray scale is $\pm10\%.$  North is $24.1^\circ$
measured counterclockwise
from the vertical axis.}
\label{fig:n6876_im}
\end{figure}

\begin{figure}[thbp]
\plotone{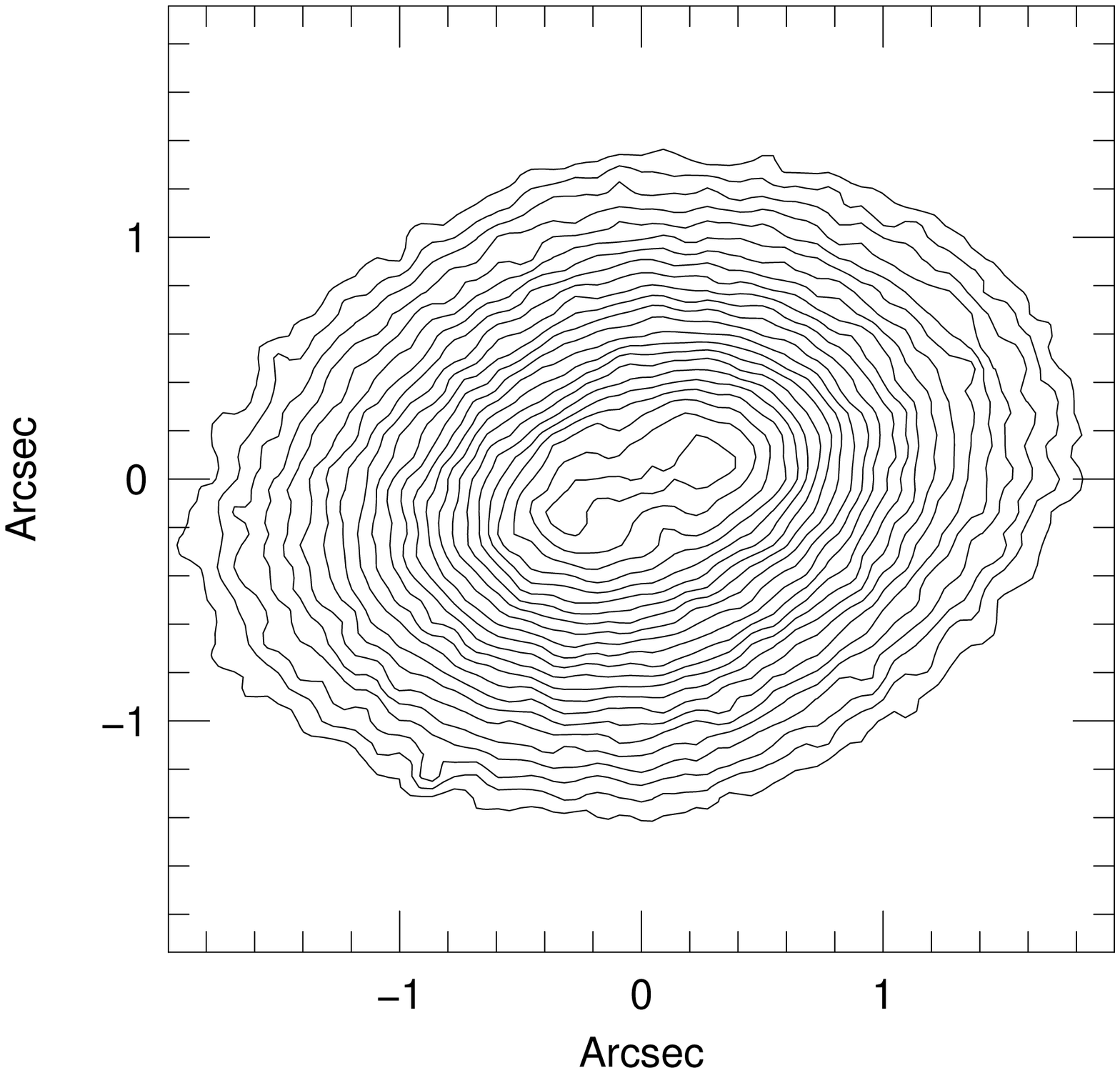}
\caption{A contour map of NGC 6876 in the deconvolved F814W image.
Contours are spaced by 0.05 mag in surface brightness; the outermost
contour corresponds to $\mu_I=16.71$ mag arcsec$^{-2}$. North is 
$24.1^\circ$
measured counterclockwise
from the vertical axis.}
\label{fig:n6876_con}
\end{figure}

NGC 6876 is a giant elliptical galaxy in the Pavo,
or \citet{f89} group 269 (4078 km s$^{-1}$ group velocity).
Its luminosity is typical for a core galaxy \citep{f97}.
Figure \ref{fig:n6876_im} shows the deconvolved
WFPC2 F555W and F814W images of NGC 6876.
The depression in surface brightness interior to the core is subtle, but is
evident as a reduced band of surface brightness along the minor
axis in both colors.
A contour map (Figure \ref{fig:n6876_con}) of the slightly smoothed
F814W image also highlights the unusual structure.
The inner isophotes of NGC 6876 become increasingly flattened,
ultimately developing an indentation on the minor axis;
NGC 6876 appears to have a
diffuse edge-on torus of starlight added to an otherwise normal core.

\begin{figure}[thbp]
\plotone{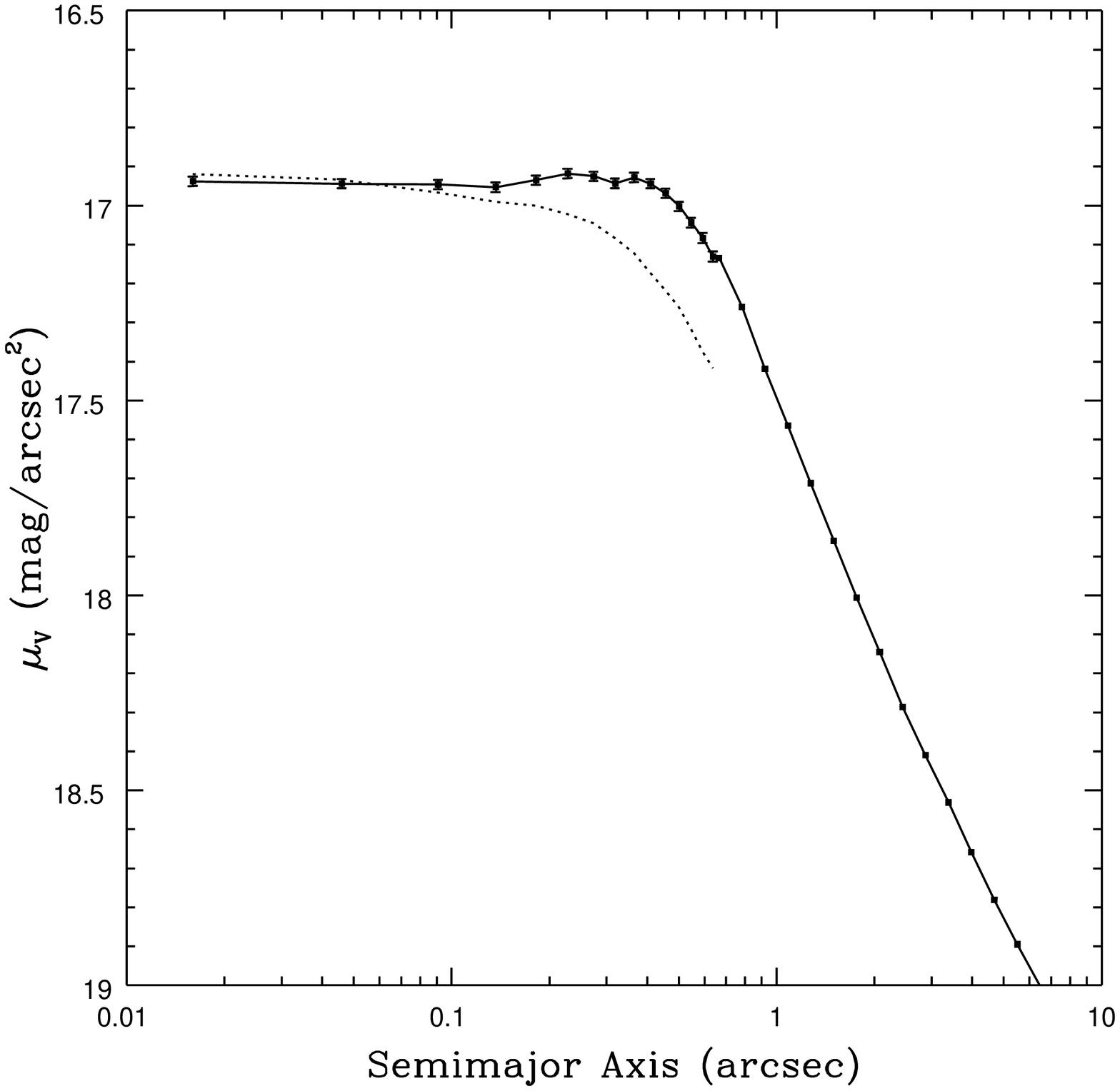}
\caption{Major axis (solid) and minor axis (dotted) deconvolved F555W
brightness profile for NGC 6876.  The profile at $r>0\asec7$
was measured by isophote fitting (connected dots).  At smaller radii the profile
for each axis was measured from a cut along the axis (of width 0\farcs14),
with the opposite sides across the nucleus averaged.}
\label{fig:n6876_cutv}
\end{figure}

Figure \ref{fig:n6876_cutv} shows the F555W major axis and minor-axis
surface brightness profiles of NGC 6876.  For $r>0\asec7,$ the profiles were
derived by the standard fitting of ellipses to the isophotes; at smaller radii,
where the isophotes were poorly fitted by ellipses, the profiles consist
simply of cuts along each axis (of 0\farcs23 width), with the opposite
sides about the nucleus averaged.
The point of maximum brightness along the major axis occurs
at $0\asec24$ (60 pc) from the nucleus,
where the surface brightness has brightened by 0.02 mag over its central value.
The minor-axis profile actually has a cusp with $\gamma$ slightly
positive over the same radii,
again suggestive of a diffuse torus that has its points of maximum projected
brightness offset along the major axis.
Looking to larger radii, it is noteworthy that the major axis
brightness profile for $1''<r<3''$ has a steeper power-law index than
that of the envelope at $r>3''.$  NGC 6876 thus has a subtle
form of a ``nuclear rise'' exterior to its nominal core.

The similar appearance of NGC 6876 in both the $V$ and $I$
filters implies that its unusual core morphology is not due to dust.
The $V-I$ map in Figure \ref{fig:n6876_im} shows no structure.
The average $V-I$ color in two $0\asec32$ boxes ($7\times7$ pixels)
centered on the maxima of the presumed torus at $0\asec31$ on either
side of the center is $1.326\pm0.005$ mag, as compared to $1.328\pm0.007$
mag for the same-sized box about the center, itself;
the net difference $\Delta(V-I)=0.002\pm0.009$ mag
going from the outer core into the center is clearly insignificant.
At the same time, since $A_V\approx2.5\Delta(V-I)$
and the central dimming is only 0.02 mag
in $V,$ it may be difficult to rule out an alternative model consisting
of weak dust absorption in a core with an exceedingly weak cusp.
As noted above, a uniformly flat core for $r<0\asec24$ would still
imply a stellar {\it density} profile that decreased at smaller radii.
For cusps steeper than even $\gamma\approx0.02,$ however, the implied
dust absorption and associated $V-I$ reddening would already be
inconsistent with the observations.

\begin{figure}[thbp]
\plotone{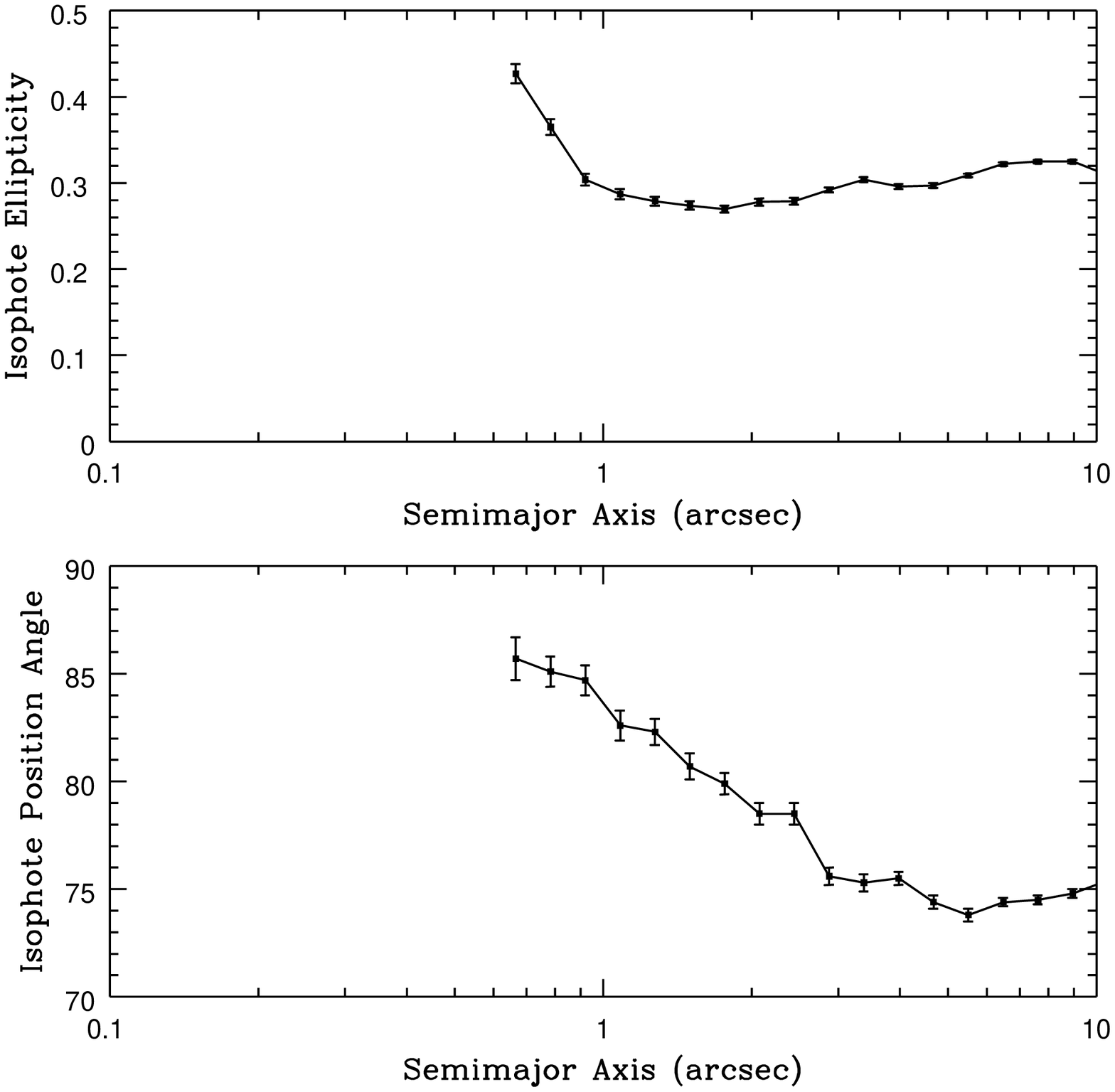}
\caption{Isophote ellipticity and position angle profiles for NGC 6876.
No points are shown for isophotes with semimajor axes $<0\asec6,$
as these isophotes are poorly fitted by ellipses.}
\label{fig:n6876_shape}
\end{figure}

\begin{figure}[thbp]
\plotone{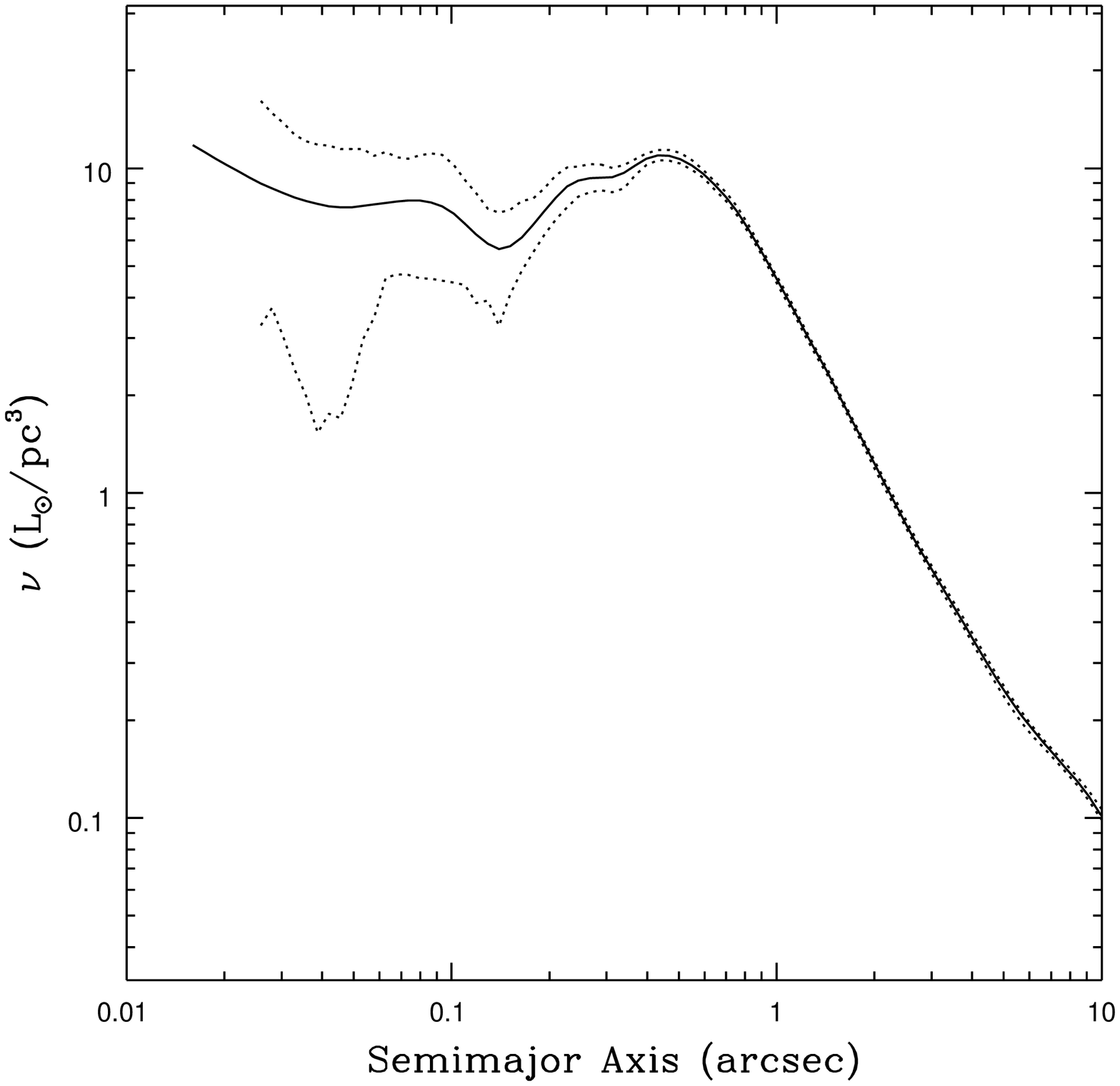}
\caption{$V$-band luminosity density profiles for the major axis
of NGC 6876.  Dashed lines give the $\pm1\sigma$ error envelopes.
The profile has been corrected for extinction.}
\label{fig:n6876_den}
\end{figure}

It is noteworthy that the PA of the innermost isophotes,
PA = 89$^\circ$, is twisted by $13^\circ$ from the major axis of the
galaxy at $r>2\asec9$ (PA = 76$^\circ$); given the high ellipticity
of the inner isophotes, this apparently modest twist is actually
highly significant.
As with NGC 3706, the isophote ellipticity and PA profiles of
NGC 6876 (Figure \ref{fig:n6876_shape}) suggest that its torus
is a more extensive system than might be inferred from the small
radius of the major axis brightness profile maximum.
The implied dip in stellar density (Figure \ref{fig:n6876_den})
is less impressive than those in NGC 3706 and NGC 4406, but still
shows a significant inward decrease.

\subsection{Brightest Cluster Galaxies}

\begin{figure}[thbp]
\plotone{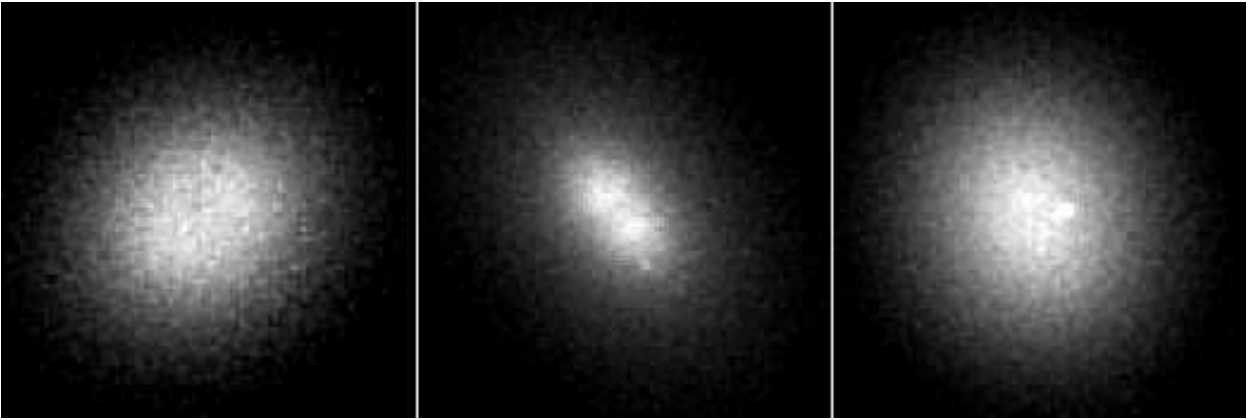}
\caption{Three BCG candidates with centrally depressed stellar
density profiles.  Left to right, the galaxies are the BCG in
A260, A347, and A3574.  Each panel is a $4'' \times 4''$ subset
of the deconvolved F814W PC snapshot image.  The intensity
stretch is arbitrary.  North is $98.6^\circ$ measured clockwise
from the vertical axis for A260, $175.1^\circ$
measured counterclockwise
from the vertical axis
for A347, and $42.4^\circ$ measured clockwise from the vertical
axis for A3574.  A faint point-source is visible in the A3574 core.}
\label{fig:bcg_im}
\end{figure}

\begin{figure}[thbp]
\plotone{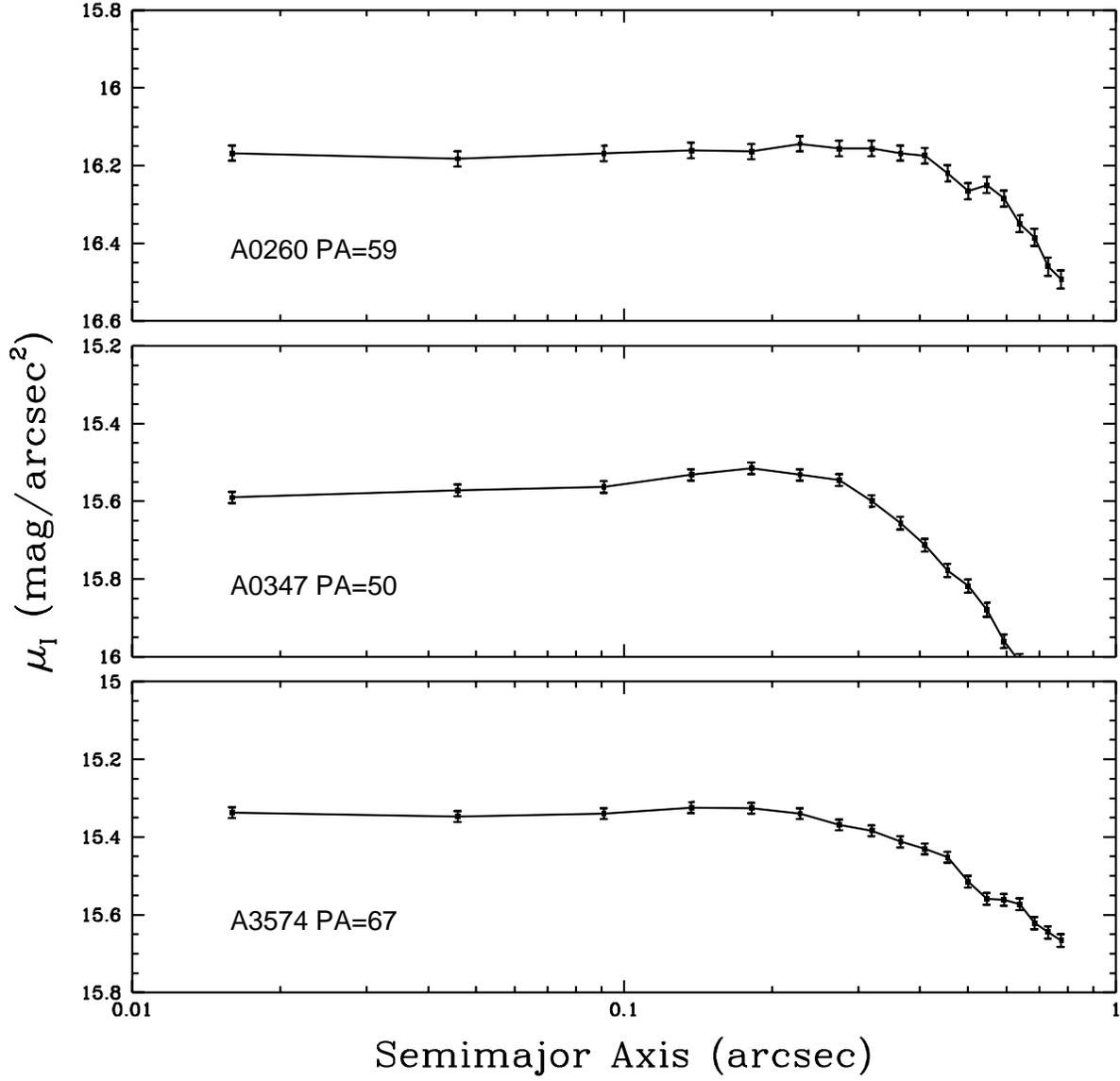}
\caption{Surface brightness profiles are shown for the
three BCG candidates with centrally depressed stellar density profiles.
The profiles were measured from cuts along the major axis of width 0\farcs23,
with the opposite sides averaged.  The position angles
of the cuts are noted in each panel.}
\label{fig:bcg_cut}
\end{figure}

A search through the {\it HST} ``snapshot'' images of BCGs 
discussed in $\S\ref{sec:obs}$ turned up a few more
candidates for galaxies with dips in their core brightness profiles.
As the snapshot images were only obtained in F814W and have relatively
low S/N, it is more difficult to rule out dust
absorption --- already clearly evident in many BCGs ---
as an explanation for such galaxies.
BCG candidates for galaxies with true local minima
in their stellar density profiles were those that most closely
resembled the three galaxies discussed above.  That is, the
minima had to be diffuse in appearance, and bracketed by bilaterally
symmetric brightness maxima.  The first criterion yielded 10 candidates
from the sample of 72 observed so far; addition of the second criterion
limited the final candidate list to three BCGs (A260, A347, and
A3574).  Images of the centers of the three galaxies are shown in
Figure \ref{fig:bcg_im} and symmetrized major axis
brightness traces are presented in Figure \ref{fig:bcg_cut}.

A347 is the best candidate for a BCG with a stellar torus.
Its isophotes become highly flattened at small radii, and its
central brightness minimum is clearly evident.  A3574 and A260
have large diffuse cores that on close examination have
subtle ($\sim 2$\%) surface brightness depressions along the major
axes interior to their break radii.
Demonstration that dust is not playing a role in all three BCGs requires
deeper exposures in at least two colors.

\section{Discussion}

It is possible that all six galaxies identified
in this paper harbor stellar tori superimposed on normal cores.
Certainly, the bright and sharply defined
feature in NGC 3706 is difficult to explain as anything else.
It is tempting to include M31 and NGC 4486B in this class,
since the \citet{trem} model for M31 is based on a stellar torus,
even though two brightness maxima (``double nuclei") in both galaxies are more
pronounced and are less symmetric than in the galaxies described here.
The origin of such tori will be discussed further below,
as well as the hypothesis that some of the ``tori'' may really
be cores that have been partially evacuated.

In either case, a genuine minimum in the density profile
implies significant rotation, triaxiality, or anisotropy.
An inward decrease in density cannot occur in spherical
galaxies whose phase-space
distribution functions depend only on the energy.  Such functions are
guaranteed to be solutions of the collisionless Boltzmann equation.
The density in this case for a spherical system is given by 
\begin{equation}
\rho=2^{3/2}\pi\int_0^\psi 
f(\epsilon)(\psi-\epsilon)^{1/2}d\epsilon,
\end{equation}
where $f\ge0$ is the mass per unit volume of phase space, 
$\epsilon\equiv -E$, and $\psi\equiv-\Phi$ (the quantities $\epsilon$
(total energy) and $\psi$ (gravitational potential) are defined so that we work
with non-negative variables: $\psi(r)$ is positive at 
all radii, and $\epsilon$ is positive for all bound stars). 
We can use 
\begin{equation}
d\rho/d\psi=2^{1/2}\pi\int_0^\psi
f(\epsilon)(\psi-\epsilon)^{-1/2}d\epsilon
\end{equation}
to evaluate the density gradient anywhere in the galaxy (note that 
the derivative with respect to the upper limit of the integral is 
zero because the integrand goes to zero).   
Since $\psi$ is a function only of radius, the gravitational field 
and any density gradient are both radial, and the latter is 
\begin{equation}
{d\rho \over dr} = {d\psi \over dr} \, {d\rho \over d\psi}.
\end{equation}
Since $d\psi/dr=-GM(r)/r^2$ is always negative and $d\rho / d\psi$ is always positive 
[$f(\epsilon)$ is non-negative for all $\epsilon$], 
$d\rho /dr$  is always negative in any spherical system
in which the distribution function depends only on energy. 
Thus, the systems observed in this paper are guaranteed to be either
nonspherical or anisotropic, and may be both.
All the systems for which we have dynamical observations indeed appear
to have significant central rotation.  The rotation amplitude in NGC 4406
is modest, but it is strong in M31, NGC 3706, and NGC 4486B.

We next discuss two possibilities for the formation of the
toroidal stellar systems.

\subsection{Star Formation and Nuclear Disks}

There are several elliptical galaxies or bulges that harbor central
stellar disks. \citet{k01} present a short list of examples and argue
that these disks were created {\it in situ} by accreted gas funneled into 
the center. \citet{svdb} emphasize that the disks can be radially
compact and of high surface brightness. 
One might then ask if the tori in the present systems result
from {\it in situ} star formation as well.  Of the six galaxies,
the relatively cold and dense ring in NGC 3706 comes closest to resembling
a stellar disk formed directly from a gas disk.  One possible scenario is
that the overall core structure was formed by the cannibalization
of a low-mass stellar system, which contained some amount of gas, by NGC 3706.
As is discussed in $\S$\ref{sec:n3706}, the ring in NGC 3706 is embedded in a
larger stellar system that itself rises above and is twisted from
the envelope brightness profile.  This system would contain the
pre-existing stars in the denser portions of the cannibalized object,
while the abrupt transition to the bright ring, itself, would reflect stars
formed by gas delivered to the center of NGC 3706 in the same merger event.
The issues remain: (1) Why has the ring not been filled in?
(2) Why has the ring and surrounding stellar system not settled
to the midplane defined by the envelope?
Presently, the maximum ring density falls well inside the Roche radius of the
estimated nuclear black hole mass.
Thus, star formation in the ring
requires either that the black hole was initially less massive, or that the
pre-existing gas was in the form of dense molecular clouds, or that the
gas layer in the disk was thinner than the current stellar disk. 

The tori in the five remaining galaxies, however, appear to be considerably
more diffuse and are less suggestive of thin stellar disks than the ring
in NGC 3706.  Further, the transition from the outer cores to the radii
of the tori appears to be smooth and gradual --- there is no abrupt transition
to a high surface brightness ring, as is seen on the minor axis of NGC 3706.
While it may be likely that these remaining tori were created in a merger
event, as well (as we argue below), there is presently no compelling evidence
that suggests that gas infall, followed by star formation within the cores,
played a role in their formation.

\subsection{Mergers of Galactic Nuclei Containing Black Holes}

It is now believed that the orbital decay of the massive black holes added to a
galaxy by mergers with other galaxies may largely determine
the inner distribution of starlight in the merger remnant.
We speculate that the unusual structures seen in the present sample (with the
possible exception of NGC 3706) may be evidence of this process.
As it happens, both the ``stellar torus added'' and
``evacuated core'' interpretations of the observations
may be supported by this scenario.

\citet{bbr} argued that galaxies harboring massive black holes should
occasionally merge and sketched out the orbital decay of the binary
black hole that would be created in such an event.  Among other predictions,
they suggested that in the later stages of the binary's life, the
two black holes would eject stars from the newly merged core as the binary
hardened, creating a local minimum in stellar density at the few parsec scale.
\citet{ebi} studied this problem numerically and showed that the binary
black hole would generate a core with a shallow cusp in the merger remnant.
Intriguingly, \citet{mak97} showed that local minima in stellar density
occurred within the core in some simulations.

The preservation of cores during mergers essentially demands that the
final stages of such events are dominated by nuclear black holes \citep{f97}.
The central density contrast between core and power-law galaxies is strong
enough that even modest cannibalism of the latter by the former should have
filled in the diffuse cores long ago.  The ``core within a core'' merger
endpoint hypothesized by \citet{k84} prior to {\it HST} observations, however,
has never been found.  The tidal field of the nuclear black hole
solves this problem by shredding the incoming nucleus prior
to its final delivery to the center.  That cores exist only in the more
luminous non-rotating elliptical galaxies, where gaseous dissipation may be
less important during the merging of their progenitors,
bolsters this picture \citep{f97}.

All six galaxies discussed here appear to be core galaxies.
The strong central isophote twist in NGC 3706 and
the kinematically decoupled core in NGC 4406 suggests that mergers
have influenced the inner structure of both galaxies.
The increase in isophote ellipticity in NGC 6876 and A347
with decreasing radius, but at radii larger than at of the
putative stellar tori, may also
suggest cannibalization of a pre-existing stellar system.
The cannibalized system would be less luminous, but denser and dynamically
colder --- the ellipticity increase would reflect stars from such
a galaxy being preferentially deposited in the core of the more luminous galaxy.

One immediate concern is why
evacuated cores are so rare if they are a natural consequence of the process
(decaying binary binary holes) that is supposed to make all cores.
It is noteworthy that the recent simulations
of \citet{mnm} only produced the $\gamma>0$
cusps seen in ``normal'' core galaxies.
The nearly constant density cores generated by \citet{mak97}, let alone
cores with dips in density, were not seen.
Significantly, \citet{mnm} always merged two power-law galaxies to
test the initial formation of a core as advocated by \citet{f97}, while
\citet{mak97} started with cores already present in the merging galaxies.
In a merging hierarchy, the more luminous galaxies may experience
multiple episodes of merging or cannibalism.  If a merger product
is used as the input for subsequent mergers, then there is a variety
of possible initial structural forms.
Two core galaxies may merge, a core galaxy may
cannibalize a power-law galaxy, and so on; allowing for the
concentration of the pre-merger cores provides an additional variable.
If the \citet{mnm} simulations are correct for the original
formation of a core galaxy, but the \citet{mak97} simulations are
correct that central evacuation can occur with appropriate initial
conditions, then perhaps the rarity of observational examples of
evacuated cores merely reflects the likelihood of these conditions.

As the black hole binary hardens, it will eject stars from the merged
core to increasingly large distances.  Early in its life, however,
when the separation of the two black holes is similar to the core
break radius, tidally stripped stars originally bound to either hole
may still linger within the core.
\citet{hbr} emphasize that under some circumstances, such as
when one galaxy has a relatively high nuclear stellar density
and arrives at the center of the other galaxy with a high impact
parameter, stars stripped from the incoming galaxy may form
a diffuse torus.  \citet{zb} have also discussed the formation of a stellar
torus as one black hole strips stars from the other.
In this picture, we really are seeing diffuse stellar
tori superimposed against the cores of the merger remnants.
Ironically, \citet{hbr} attempted to explain the thin nuclear disks
(as discussed in the context of NGC 3706) as being created by this
process, but noted that they could only generate diffuse tori,
not the high aspect-ratio disks seen in systems like NGC 3115 or NGC 4594.
Some of the present systems may be the realizations of
the toroidal structures seen in the \citet{hbr} simulations.

\section{Summary}

We have discussed six early-type galaxies
in which the luminosity density distribution in the core
cannot be represented as a simple power law.
The surface brightness profiles of all six systems
decrease over limited radii interior to their cores.  This
implies that there are correspondingly more pronounced dips in their
luminosity density profiles.  In the systems for which we
have multi-color imaging (NGC 4406 and NGC 6876),
it is unlikely that we are being
confused by some low level of diffuse dust absorption.
The three BCG strongly resemble these two systems, and they show
no compact dust clouds anywhere around their cores.
We have therefore included them within this discussion,
although additional multicolor imaging of the BCG
should be obtained to verify this conclusion.
The final system, NGC 3706, harbors a bright, apparently edge-on stellar ring.

There are two literally complementary interpretations of the core
structures seen in the six galaxies.  The first picture is that
we are seeing stellar tori superimposed on otherwise
normal cores.  The narrow ring in NGC 3706 stands in such strong
contrast with its surrounding core that it is difficult to accept
any alternative interpretation.  NGC 3706, and perhaps
some of the five remaining systems, may be related to the double-nucleus
galaxies M31 and NGC 4496B. 
The six galaxies discussed here did not present themselves
as double nuclei; however, as they all do have subtle local maxima
in surface brightness that bracket their apparent nuclear centers,
it would take only a modest brightness enhancement at these locations to have
them appear as doubles.  Instead, perhaps the more suggestive
presence of stellar tori in the present galaxies bolsters the
plausibility of this interpretation for M31 and NGC 4486B, as well.

The thinness of the NGC 3706 torus probably requires it to have formed
dissipatively.  The core structure of NGC 3706 suggests that it and the inner
ring were formed in a merger or cannibalism of a stellar system
that contained some amount of gas.  In the ``torus-added''
picture the tori in the remaining systems may have formed
from the cannibalization of gas-free stellar systems.
In this case, the tori would have been generated by the tidal
disruption of the cores or central cusps of the original systems
by a massive black-hole binary formed from the nuclear black holes
contained in the original systems.

In the second picture we are seeing a true depletion
of stars within the cores of the galaxies.  As in the picture above,
a black hole binary would be formed in a merger.  As the binary
hardens, it ejects stars from the core.  The difference between
this picture and the ``torus-added'' scenario may be one of initial
conditions, in which ejection of stars at later stages in the binary's
evolution competes against deposition of relatively colder stars
around the core at an earlier stage in the binary's life for domination
of the final integrated structure of the core.

The ``core-evacuated'' scenario is suggested by theoretical results that are
only weakly evident in some merger simulations.  If
core evacuation does occur, it may require rare initial conditions.
The \citet{mnm} simulations argue that evacuation is not likely
to be seen in the merger of two power-law galaxies that would create
a core galaxy initially.  It may be that evacuated cores can occur
in subsequent mergers in which at least one of the
two galaxies already contains a core.

Regardless of which scenario is correct, we present the six galaxies
as interesting probes on the formation of cores.
Theoretical and observational work both point to a strong role
for nuclear black holes for establishing the core structure of
merger remnants.  Rare and unusual core structures may offer a unique
narrative in this story not otherwise voiced by more normal galaxies.

\acknowledgments

We thank Joseph Jensen for providing the reduced NIC2 images of
NGC 4406.  Support for GO proposals 5512, 6099, 6587, and 8683 was provided by
NASA through grants from the Space Telescope Science Institute, which is
operated by the Association of Universities for Research in Astronomy, Inc.,
under NASA contract NAS 5-26555.

\clearpage

\begin{deluxetable}{ccccl}
\tablecolumns{5}
\tablewidth{0pt}
\tablecaption{Observational Summary}
\tablehead{\colhead{Galaxy}&\colhead{D (Mpc)}
&\colhead{$M_B$ (mag)}&\colhead{$A_B$ (mag)}&\colhead{Images}}
\startdata
NGC 3706&35&$-$21.24&0.36&500 s F555W ($2\times2$ subsampled) \\
NGC 4406&16&$-$21.26&0.11&500 s F555W, 500 s F814W, 140 s F160W \\
NGC 6876&51&$-$22.43&0.16&4100 s F555W, 4100 s F814W \\
\enddata
\label{tab:obs}
\tablecomments{$H_0=80~{\rm km~s^{-1}~Mpc^{-1}}$ is assumed.
Group velocities were used for distance estimation.}
\end{deluxetable}

\clearpage

\end{document}